\documentclass[doublecol]{epl2} 
\usepackage{amsmath}
\usepackage{graphicx}

\title{Stimulated Raman Scattering and Molecular Modulation in Anti-resonant Hollow-core Fibres}
\shorttitle{SRS and Molecular Modulation in ARFs} 

\author{P. Arcos\inst{1}\thanks{These authors contributed equally to this work.} \and A. Mena\inst{1}\footnotemark[1] \and M. S\'anchez-Hern\'andez\inst{1}\footnotemark[1] \and E. Arrospide\inst{2} \and G. Aldabaldetreku\inst{1} \and M. A. Illarramendi\inst{3} \and J. Zubia\inst{1,4} \and D. Novoa\inst{1,4,5}\thanks{E-mail: \email{david.novoa@ehu.eus} (corresponding author)}}
\shortauthor{P. Arcos \etal}

\institute{                    
  \inst{1} Department of Communications Engineering, University of the Basque Country (UPV/EHU) - Torres Quevedo 1, 48013 Bilbao, Spain\\
  \inst{2} Department of Applied Mathematics, University of the Basque Country (UPV/EHU) - Torres Quevedo 1, 48013 Bilbao, Spain\\
  \inst{3} Department of Applied Physics, University of the Basque Country (UPV/EHU) - Torres Quevedo 1, 48013 Bilbao, Spain\\
  \inst{4} EHU Quantum Center, University of the Basque Country (UPV/EHU) - 48013 Bilbao, Spain\\
  \inst{5} IKERBASQUE, Basque Foundation for Science - Plaza Euskadi 5, 48009 Bilbao, Spain
}

\abstract{
Raman scattering is the inelastic process where photons bounce off molecules, losing energy and becoming red-shifted. This weak effect is unique to each molecular species, making it an essential tool in e.g. spectroscopy and label-free microscopy. The invention of the laser enabled a regime of stimulated Raman scattering (SRS), where the efficiency is greatly increased by inducing coherent molecular oscillations. However, this phenomenon required high intensities due to the limited interaction volumes, and this limitation was overcome by the emergence of anti-resonant fibres (ARFs) guiding light in a small hollow channel over long distances. Based on their unique properties, this Perspective reviews the transformative impact of ARFs on modern SRS-based applications ranging from development of light sources and convertors for spectroscopy and materials science, to quantum technologies for the future quantum networks, providing insights into future trends and the expanding horizons of the field.}

\begin{document}

\maketitle

\section{Introduction}

As discovered about a century ago \cite{raman1928new}, photons can spontaneously scatter off molecular ensembles, losing energy in the process and therefore becoming red-shifted in wavelength compared to the incident radiation. This inelastic optical phenomenon named Raman scattering after its discoverer, revolutionised spectroscopy since the difference between the incident (Pump) and scattered (Stokes) photon frequencies – the Raman shift $\Omega_R$ – is a unique fingerprint of the interrogated molecular species. However, the probability of this inelastic scattering process is typically very low (in diatomic gases like H$_2$, typically only 1 in 10$^9$ photons will be red-shifted), making the detection of the spontaneous Raman signals challenging under moderate illumination conditions. In this regard, the invention of the laser \cite{maiman1960stimulated} marked a turning point as the high intensities attainable using coherent light unlocked a new regime of stimulated Raman scattering (SRS) above a certain threshold \cite{SRS1962}. In SRS the intense pump light and the spontaneously generated Stokes photons induce synchronous molecular oscillations in the medium \cite{RaymerWalmsley1985, HarrisSolokov1998} that boost exponential amplification of the Stokes band up to levels comparable to the initial pump, creating more “optical phonons” in the process. These molecular coherence waves can also facilitate the up-conversion of the pump by $\Omega_R$ to the anti-Stokes band. Notably, they also enable the modulation of arbitrary optical signals injected in the Raman-active medium after the molecular coherence has been established \cite{nazarkin1999generation, bauerschmidt2014supercontinuum, mixing_2015}. Such molecular modulation process can be very efficient provided the inherent linear momentum of the coherence waves is matched to that of the corresponding optical transitions of interest.
\begin{figure*}[ht!]
\begin{center}
\includegraphics{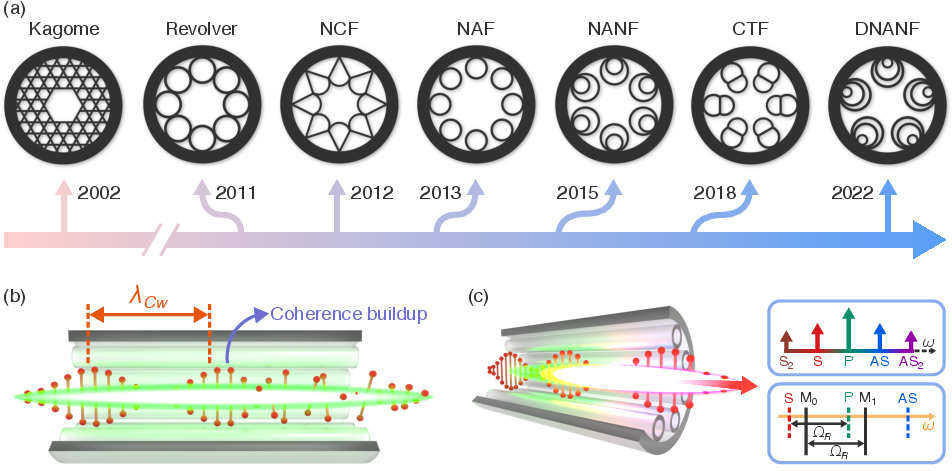}
\caption{(a) Overview of the evolution of ARFs with various design milestones: kagome \cite{2002_Bernabid}, revolver \cite{vincetti2010waveguiding, 2011_Revolver_Moscow}, negative-curvature fibre (NCF) \cite{Yu12}, nodeless anti-resonant fibre (NAF) \cite{2013_NAF_Moscow}, nested anti-resonant nodeless fibre (NANF) \cite{poletti2014nested, 2015_NANF_Southampton}, conjoined tube fibre (CTF) \cite{2018_CTF_Beijing} and double-nested anti-resonant nodeless fibre (DNANF) \cite{2024_DNANF_Southampton}. (b) Excitation of Raman coherence waves of wavelength $\lambda_{CW}$ in ARFs. (c) Molecular modulation yielding the generation of Raman combs (top-right panel) or thresholdless frequency up-conversion of a mixing beam M$_0$ to its anti-Stokes band M$_1$ (bottom-right panel). $\Omega_R$ is the Raman transition frequency and P, S, S$_2$, AS and AS$_2$ represent the pump and the different Stokes and anti-Stokes bands.}
\label{fig:intro-figure}
\end{center}
\end{figure*}
Despite its significance and diverse applications in spectroscopy \cite{SRSspectroscopy}, microscopy \cite{SRSmicroscopy} and laser science \cite{SRSlasers}, the high threshold for SRS dynamics in common media like molecular gases still requires the use of powerful lasers to compensate for a limited interaction volume in free-space geometries, which has hampered progress in these areas.

The advent of broadband-guiding anti-resonant hollow-core fibres (ARFs) filled with gases offered a promising route to overcome this limitation. ARFs confine both light and matter in a small hollow channel over long distances \cite{russell2014hollow}, drastically reducing the SRS threshold by orders of magnitude compared to free-space arrangements \cite{2002_Bernabid} and enabling high performance with simpler pump lasers. Initially conceived as broadband, reduced-latency alternatives to standard solid-core fibres for optical communications, ARFs guide light through anti-resonant reflection \cite{antiresonanttheory} and inhibited coupling between core-guided light and the surrounding cladding elements \cite{Raman2007_0.325_Bernabid}. These mechanisms allow for ultralow-loss broad transmission windows, tight light-matter confinement in the core and resilience to photo-induced damage. Since the seminal kagome-style design \cite{2002_Bernabid}, ARFs have undergone numerous structural changes (see e.g. Fig. \ref{fig:intro-figure}a and Refs. \cite{vincetti2010waveguiding, 2011_Revolver_Moscow,Yu12,2013_NAF_Moscow,poletti2014nested, 2015_NANF_Southampton,2018_CTF_Beijing,2024_DNANF_Southampton}) aiming at both reducing their overall attenuation and tailoring other optical properties \cite{modalcontent,2018_CTF_Beijing,helicalSRPCF_2018}. Nowadays the most widespread version of ARF comprises a single tubular lattice of contact-free hollow capillaries surrounding the core (“NAF” design in Fig. \ref{fig:intro-figure}a). Remarkably, the losses can be dramatically reduced by nesting more capillaries within each cladding unit, making ARFs the most transparent optical waveguides ever developed \cite{Jasion20,2024_DNANF_Southampton}. In addition, gas-filled ARFs offer a pressure-tunable dispersion landscape that enables additional control over strong light-matter interactions along the fibre length \cite{mixing_2015}. This is particularly advantageous for applications involving SRS-based molecular modulation (see panels (b)-(c) in Fig. \ref{fig:intro-figure}), as it allows fine adjustment of momentum mismatch among the interacting fields over a broad bandwidth. This platform not only surpasses free-space systems for the applications mentioned above, but also opens new routes in emerging fields such as quantum communications.

In this brief Perspective we revisit the state-of-the-art applications of SRS and molecular modulation in gas-filled ARFs. We begin with a concise theoretical introduction of the modelling of in-fibre SRS dynamics. We then review recent applications of this platform in key transversal areas such as light-source development, multidimensional frequency conversion or quantum technologies. Finally we provide our perspective on future trends in the field and the horizons that lie ahead.

\section{Theory and modelling}

The dynamics of a pump pulse of duration comparable to the molecular coherence lifetime $T_2$ (i.e. in the so-called ``transient'' regime\cite{chen2024unified}) travelling through an ARF filled with a Raman-active gas is governed by the following Maxwell-Bloch equations \cite{gainsup}:
\begin{eqnarray}
    \frac{\partial E_{\sigma,l}}{\partial z} \hspace{-2ex}&=&\hspace{-2ex} - i\sum_{\nu\chi\eta}^M\frac{s_{\sigma\nu\chi\eta}}{s_{\nu\chi}}\bigg( \kappa_{2,l}\frac{\omega_l}{\omega_{l-1}} Q_{\nu\chi} E_{\eta,l-1}q_{\eta,l-1}q_{\sigma,l}^*  \nonumber \\ && \hspace{-2ex} + \kappa_{2,l+1}Q_{\nu\chi}^*E_{\eta,l+1}q_{\eta,l+1}q_{\sigma,l}^* \bigg) - \frac{1}{2}\alpha_{\sigma,l}E_{\sigma,l}, \hspace{2ex} \label{eq:eqsMB}\\
     \frac{\partial Q_{\nu\chi}}{\partial t} \hspace{-2ex}&=&\hspace{-2ex} -\frac{Q_{\nu\chi}}{T_2} - \frac{i}{4}s_{\nu\chi}\sum_l^M\kappa_{1,l}E_{\nu,l}E_{\chi,l-1}^*q_{\nu,l}q_{\chi,l-1}^*, \hspace{2ex}
     \label{eq:MBeqs}
\end{eqnarray}
 where the integer $l$ denotes the sideband of frequency $\omega_l = \omega_p + l\Omega_R$ ($\omega_p$ is the central frequency of the pump) and the summations go over all possible permutations of the modal set M that comprises the linearly-polarised-like core modes supported by the fibre. $\kappa_{1,l}$ and $\kappa_{2,l}$ are the coupling constants, $\alpha_{\sigma,l}$ are the modal attenuation coefficients \cite{vincetti2019simple}, and $q_{\sigma,l} = \exp(-i\beta_{\sigma,l}z)$, where $\beta_{\sigma,l}$ is the pressure-dependent propagation constant of the band $l$ in spatial mode $\sigma$\cite{zeisberger2017analytic,russell2014hollow}. {\it Q} and {\it E} are the molecular coherence amplitude and electric-field envelope, respectively, and $s_{ijkl}$ represent the nonlinear spatial overlap integrals among different interacting fibre modes. We assume that the majority of the molecules remain in the ground state, a good approximation in most relevant scenarios. Furthermore, in order to keep our analysis simple and focused, we will consider SRS as the dominant effect in all dynamics discussed in this Perspective, although other nonlinear phenomena such as e.g. self-phase modulation, four-wave mixing and photoionisation might also be relevant in systems pumped by ultrashort pulses \cite{russell2014hollow}.

\begin{figure}[ht]
    \centering
\onefigure[width = \columnwidth]{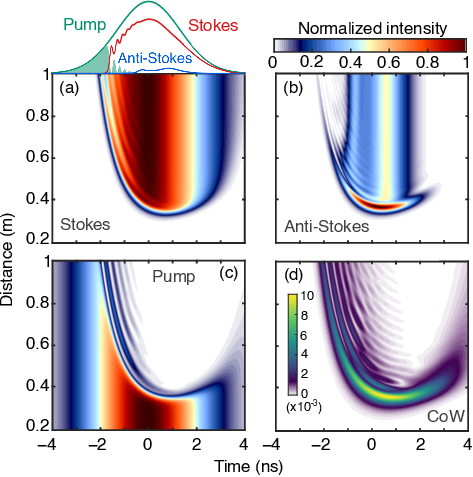}
    \caption{(a,b,c) Spatiotemporal evolution of the Stokes, anti-Stokes and pump signals along a H$_2$-filled NAF-type ARF with 20 $\mu$m core diameter, 100 nm capillary-wall thickness and 21 bar of filling pressure. The energy of the 3.8 ns-long pump pulse is 5 $\mu$J and all signals are normalised to their respective maxima. (a-top) Lineouts of the outcoupled temporal profiles of the bands normalised to the initial pump value (green line). The green shaded area shows the pump depletion. (d) Spatiotemporal evolution of the Raman coherence amplitude. Time is measured relative to a reference frame co-moving at the group velocity of the pump.}
    \label{fig:propagation}
\end{figure}

\begin{figure}
    \centering
\onefigure[width = \columnwidth]{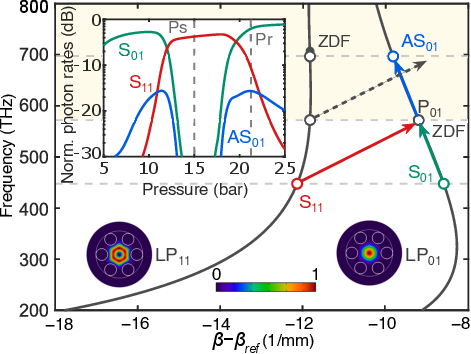}
    \caption{Dispersion diagram of the ARF described in Fig. \ref{fig:propagation} when filled with 15 bar H$_2$. The dispersion curves of the LP$_{01}$- and LP$_{11}$-like core modes with their simulated intensity profiles are plotted against $\beta - \beta_{ref}$, where $\beta_{ref}$ is a linear function of frequency used to bring out the S-shaped profile. Colour-coded arrows represent intramodal (green) and intermodal (red) coherence waves, as well as intramodal (blue) and intermodal (black-dashed) pump-anti-Stokes transitions. ZDF stands for zero-dispersion frequency and the shaded region indicates normal dispersion for the LP$_{01}$-like mode. Inset: Outcoupled photon rates (in decibel scale) of the pump (P), Stokes (S) and anti-Stokes (AS) bands for different pressures. The subscripts indicate the mode profile of the band and P$_r=21$ bar. At P$_S$ $\approx$ 15 bar the rates of intramodal phonon creation and annihilation precisely balance resulting in total suppression of the intramodal gain and enhancement of intermodal interactions.}
    \label{fig:s_shape}
\end{figure}

To illustrate the complex SRS-driven dynamics, in Fig. \ref{fig:propagation} we display the simulated evolution of the pump, Stokes and anti-Stokes bands (along with the generated coherence) inside a hydrogen-filled ARF. The peak of the Stokes band (Fig. \ref{fig:propagation}a) emerges delayed with respect to that of the pump (Fig. \ref{fig:propagation}c), as expected for purely transient dynamics. On the other hand, the anti-Stokes band appears even further delayed (Fig. \ref{fig:propagation}b) since its efficient generation and amplification relies on the presence of both significant Raman coherence (Fig. \ref{fig:propagation}d) and pump signal. Phase matching between the coherence waves and the pump-to-anti-Stokes transition is also crucial for efficient anti-Stokes generation via molecular modulation. As it can be obtained from Eqs.~\eqref{eq:eqsMB} and \eqref{eq:MBeqs}, this condition can be fulfilled if the overall dephasing rate  $\vartheta = |\Delta \beta_{CoW} - \Delta \beta_M| \approx 0$, where the propagation constant of the coherence waves is given by $\Delta\beta_{CoW} = \beta_{pump} - \beta_{Stokes}$  and $\Delta \beta_M$ is the beat-note created by the corresponding transition of interest. Perfect collinear phase matching can be uniquely achieved in gas-filled ARFs owing to their pressure-adjustable S-shaped dispersion profile (see Fig. \ref{fig:s_shape}).

\begin{figure*}
\centering
\includegraphics[width=0.92\textwidth]{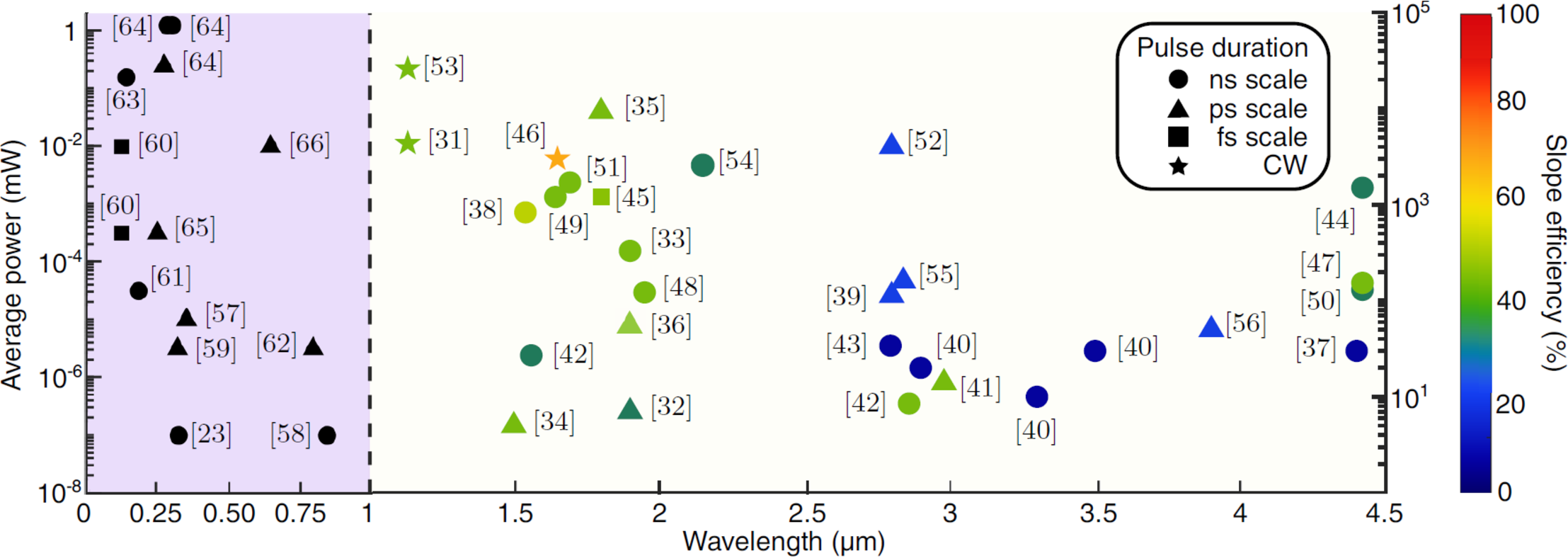}
\caption{Achievements in ARF-based Raman lasers regarding emission parameters. The different results are divided in two regions: (Left) In the UV and visible range the emitted average power is lower and with small slope efficiencies, while in the (Right) IR range these systems offer a promising alternative to reach longer wavelengths. The shapes of the data points indicate the pulse-duration scale and colour in the right diagram is related to the slope efficiency in the IR. Note that the data points in the purple-shaded area are all black given their low conversion efficiency.}
\label{fig:laser_performance}
\end{figure*}

\section{New laser sources and transport}

A deeper understanding of the underlying guidance mechanisms in ARFs sparked renewed interest in the development of new broadband fibre designs exhibiting ultralow losses in the dB/km range and below \cite{2018_CTF_Beijing, 2024_DNANF_Southampton, 2022_ReducedRoughnessUV_Limoges}.  These remarkable optical properties paved the way for a new light transport approach specially suited for ultrashort pulses since robust and flexible beam delivery based on standard solid-core fibres is severely impaired by the high-peak powers of the propagating pulses and chromatic dispersion of the waveguides, which can significantly alter the pulse shape and even damage the guiding medium. In sharp contrast, ARFs minimise these effects owing to the very small overlap of the core-guided light with the surrounding glass microstructure and weakly anomalous dispersion (if evacuated). Furthermore, the use of monoatomic filling gases or vacuum allows for the suppression of nonlinear phenomena (like SRS) in molecular gases and mixtures such as ambient air improving stability and modal beam quality \cite{2023_flexible}. As a result, remarkable beam delivery results in spectrally distant regions from UV to mid-IR have been achieved for high-power picosecond \cite{2018_Single-modeUV_Bath, 2023_20WDeliveryUV_Limoges} and femtosecond \cite{2018_HighPowerMidIR_Beijing, 2018_Single-modeUV_Bath} lasers within short distances in the metre-scale. 

Regarding beam quality, recent work has shown laser delivery with M$^2$ coefficients close to unity \cite{2022_m2, 2024_LongDistance&LifespaningUV_Limoges}. Nevertheless, maintaining beam quality over longer distances remains a challenge due to losses and pressure differences over the fibre length. Picosecond IR delivery has been achieved in a hundred-meter scale \cite{2022_100mkWPower_Southampton}, while kW-power continuous-wave transport has reached the km-scale\cite{2022_Kilowatt_kmScale_Southampton,2023_2.2kW_CREOL}. Despite this steady progress, long-distance delivery in the visible–UV remains a challenge \cite{2024_NonLinearity_Southampton}.

On the other hand, the unique properties of ARFs has straightforwardly placed them as platforms for new laser sources. Although population inversion techniques with gas-filled ARFs as laser active media have gained significant attention \cite{NonRaman2022_MidInfrared_JinanUniv, NonRaman2024_MidInfrared_ZefengWang}, SRS in gases offers an efficient way for obtaining narrowband and tunable Stokes and anti-Stokes laser emission in single-pass configuration \cite{Raman2007_1.135cw_Bath}. Overall laser performance achieved so far using different gases, pressures and fibre designs can be classified in two spectral ranges, above and below 1 $\mu$m (Fig. \ref{fig:laser_performance}). The former has experienced a huge development reaching the mid-IR up to 4.42 $\mu$m \cite{Raman2019_4.42p_Moscow, Raman2020_4.42p_DTU&CREOL} with $\sim$ 45\% efficiency, a spectral range otherwise difficult to access. Slope efficiencies raise up to 60\% for shorter IR wavelengths \cite{Raman2020_1.65cw} and drop as the Stokes emission wavelength increases. In the short wavelength edge, emission down to the vacuum ultraviolet ($\lambda<$ 200 nm) was demonstrated either via supercontinuum generation \cite{Raman2015_0.124p_MaxPlanck} (shortest $\lambda$ $\sim$ 124 nm) or through cascaded multi-order anti-Stokes emission\cite{Raman2020_0.141_MaxPlanck} (shortest $\lambda$ $\sim$ 142 nm). Nevertheless, emission in the UV range exhibits very low conversion efficiency with emitted average powers in the mW-scale for Stokes emission or one order of magnitude lower for anti-Stokes conversion \cite{Raman2022_0.274&0.289&0.299p_Limoges}.

Although some studies have achieved significant CW conversion \cite{Raman2007_1.135cw_Bath, Raman2020_1.65cw, Raman2022_1.135cw}, this technique presents a higher power threshold compared to population inversion, and therefore most of the systems developed so far are pulsed. Most studies have been carried out in the ns and ps-scale, but the femtosecond regime has recently attracted increased attention \cite{Raman2015_0.124p_MaxPlanck, Raman2020_fsBufetov, Raman2020_1.8p_MaxPlanck, Raman2020_fsBufetov2}. To conclude this section, the current efforts in the field are being directed at enhancing the UV performance of Raman lasers, as well as for extending the long-wavelength edge with prospects to even reach the far-infrared\cite{Raman2023_12p_CornellTheor}.

\section{Multidimensional conversion}

As we have discussed above, the coherence waves excited in the gaseous core of ARFs are the essential building blocks to any light-matter interaction dynamics triggered in the system such as, for example, the generation of anti-Stokes radiation. During this process, optical phonons are annihilated, whereas further Stokes emission is accompanied by the generation and amplification of coherence waves. Interestingly, as briefly explained in the theory section, when the rates of phonon creation and annihilation are precisely balanced, the Raman gain is parametrically suppressed \cite{gainsup} as shown in Fig. \ref{fig:s_shape}. This is usually the case at dephasing rates $\vartheta\approx 0$ for specific intramodal transitions, which in turn enhances otherwise weaker intermodal interactions (i.e., the conversion between pump photons in one spatial core mode to Stokes photons in another mode) that are more resilient to this effect. In this situation, not only the frequency of the light varies, but also the spatial profile of the resulting emission changes. Since the overall dephasing is pressure-adjustable, it implies that the gain-suppression regime can be accessed in a controllable way as illustrated in the pressure scan of the inset of Fig. \ref{fig:s_shape} (perfect phase matching is achieved at P$_S$ $\approx$ 15 bar in the example displayed). In addition, since dispersion is crucial to harness coherent gain reduction, using different gases and mixtures adds an extra knob to control the in-fibre nonlinear dynamics \cite{mixtures_control_transient_state,gasmixtures_shirmohammad2023collision}. Moreover, efficient SRS conversion has been observed using both vibrational \cite{Raman2016_1.5p} and rotational \cite{conv_polarisation} transitions in different gases \cite{Raman2019_4.42p_Moscow, Raman2018_2.9&3.3&3.5_Moscow,Raman2022_2.15,Raman2021_1.95p_DTU&CREOL,safaei2020high,Raman2018_2.8p_Beijing,Raman2018_1.54p,Raman2018_2.98p,Raman2019_2.796-2.863p}.

On the other hand, if the Stokes and anti-Stokes signals become intense enough, they can originate their own Raman bands (see Fig.~\ref{fig:intro-figure}(c), top-right panel), cascading into a frequency comb of spectrally equidistant signals \cite{Raman2022_0.250p_Beijing,Raman2007_0.325_Bernabid,Raman2020_0.141_MaxPlanck}. Moreover, if other quasi-instantaneous nonlinear responses are simultaneously excited in the fibre core, the spectrum of each signal can widen upon propagation, filling the gaps between bands and leading to a supercontinuum-type spectrum \cite{Raman2015_0.124p_MaxPlanck,Raman2022_0.250p_Beijing,Raman2022_0.650_Moscow}. Moreover, it has also been demonstrated that external arbitrary signals can interact with the existing Raman coherence and be modulated without threshold as long as the appropriate phase-matching conditions are fulfilled \cite{mixing_UV_threshholdless,mixing_2015,D2_mixing}.

Finally, in the context of angular-momentum control via SRS, it has been demonstrated that helically-twisted ARFs \cite{helicalSRPCF_2018} support non-degenerate circularly-polarised core modes in contrast to conventional longitudinally-invariant ARFs. This geometric property allows for the generation, control and delivery of circularly-polarised Raman bands and vortices \cite{conv_polarisation,conv_vortices}, opening the door to a new class of chiral fibre-based devices with potential applications in biospectroscopy and sensing.

\section{Lightwave quantum technologies}

\begin{figure}
    \centering
\onefigure[width = \columnwidth]{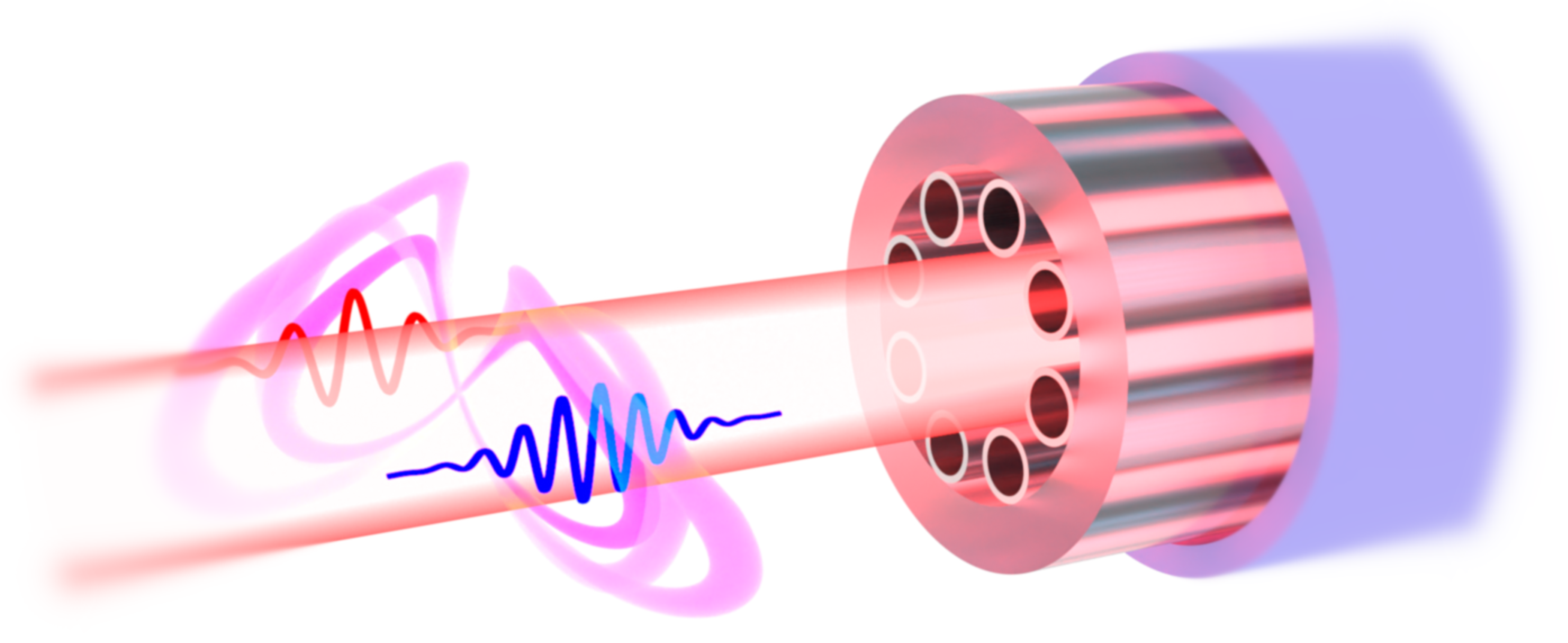}
    \caption{Artistic impression of the simultaneous transport of entangled photon pairs and high-power laser beams using an ARF. Classical and quantum communication channels can coexist in these fibres.}
    \label{fig:quantum-art}
\end{figure}

The use of ARFs as core elements to develop new light sources, transport lines and multidimensional converters has also been accompanied by a thriving interest for applications in quantum technologies. The low losses, reduced latency and broad transmission windows achieved in ARFs permit delivery of single photons over long distances with high efficiency and remarkable spatial quality. Furthermore, it has been experimentally demonstrated that ARFs preserve the quantum properties of the photons during transportation \cite{loffler2011, Chen21, antesberger2023}. These features have generated interest for their use in quantum communications, as they also allow the coexistence of classical and quantum channels \cite{obada2022, florian2023} (see Fig. \ref{fig:quantum-art} for an illustration), enabling the implementation of quantum communication protocols.

Regarding their use for quantum communications, ARFs are also an excellent platform for quantum frequency conversion as recently demonstrated in the context of correlation-preserving molecular modulation of single photons without threshold \cite{tyumenev2022}. As the efficiency of this quantum process is already close to unity, this unlocks many possibilities in, for example, the efficient interfacing between devices working at different frequencies, which is key in this communications paradigm. The use of ARFs as multidimensional converters is a promising asset for tuning different light sources.

The excellent modal quality and tight confinement in ARFs mitigates the influence of parasitic Raman effects due to undesired interactions with the cladding material. Consequently, when ARFs are filled with noble gases instead of Raman-active media, this platform can also be used to expand the catalogue of quantum light sources. For example, frequency-entangled photon pairs with large spectral separation can be directly produced via four-wave mixing \cite{Cordier20, lopez2021, lopez2023}. Additionally, nonclassical highly-correlated many-photon ``twin'' beams can be also be generated through modulational instability \cite{finger2015, finger2017}.

The central hollow channel of ARFs has also been employed to build systems for atom guidance, trapping and manipulation\cite{epple2014, gebert2014, veit2016, langbecker2017, shoichi2019, mingjie2019, leong2020, krehlik2022}, enabling applications in atomic interferometry\cite{xin2018} and spectroscopy\cite{okaba2014}. One of these possibilities is the development of quantum memories in ARFs by manipulating the atoms confined in the hollow core\cite{sprague2014, mcgarry2024, rowland2024}. Among the advantages of implementing quantum memories with this platform, we find better memory-fibre coupling efficiencies\cite{mcgarry2024} and lower power requirements\cite{rowland2024}, offering synchronisation improvements in quantum information processing.

\section{Conclusions and Perspectives}

In this Perspective we have revisited the state-of-the-art applications of stimulated Raman scattering and molecular modulation in gas-filled anti-resonant fibres. These unique systems have lifted most of the limitations of their free-space counterparts while enabling new applications in light-source development, frequency conversion and quantum technologies, whose advancement has concentrated most of the research carried out in this SRS platform over the last years.

Owing to the low light-glass overlap in ARFs, there are prospects for new ARF-based Raman lasers operating in the mid- to far-infrared. On the other side of the spectrum, improvements in their UV characteristics such as light transmission and damage threshold could yield a new approach for tunable UV laser sources. Since confinement loss in ARFs has already surpassed that of the best solid-core fibres, the next frontier is to improve the inner surface roughness through novel manufacturing techniques.
Regarding frequency conversion, we have discussed how harnessing the molecular coherence waves enabled the manipulation of the different degrees of freedom of the guided light fields. In this area, control over the total angular momentum of the outcoupled radiation is still a challenge that requires further development of chiral guiding structures.

In the field on quantum technologies, the full potential of ARFs to provide tunable quantum transducers, exotic quantum light sources or better fibre-compatible quantum memories is yet to be unleashed. Future research directions might involve improving ARF interconnectivity for quantum networking, as well as further theoretical developments to understand the preservation of quantum light properties upon molecular modulation.

\acknowledgments
Acknowledgements: This work was supported by the grants PID2021-123131NA-I00, PID2021-122505OBC31, PRE2022-102843 and TED2021-129959B-C21, funded by MICIU/AEI/10.13039/501100011033, by “ERDF a way of making Europe”, by the “European Union NextGenerationEU/PRTR” and "ESF+", and the Gobierno Vasco/Eusko Jaurlaritza (IT1452-22), ELKARTEK ($\mu$4Smart-KK-2023/00016 and Ekohegaz II-KK-2023/00051), and the “Translight” initiative of UPV/EHU, and the IKUR Strategy of the Department of Education of the Basque Government through the grant IKUR\_IKA\_23/03. M. S.-H. acknowledges support from the predoctoral grant "Formación de Profesorado Universitario" FPU22/01451 from the Spanish Ministry of Science, Innovation and Universities (MICIU). P. A. acknowledges support from the Basque Government grant for predoctoral researchers, Ref. PREGV23/47.\\


\end{document}